\documentclass{article}

\usepackage{arxiv}

\usepackage[utf8]{inputenc} % allow utf-8 input
\usepackage[T1]{fontenc}    % use 8-bit T1 fonts
\usepackage{hyperref}       % hyperlinks
\usepackage{url}            % simple URL typesetting
\usepackage{booktabs}       % professional-quality tables
\usepackage{amsfonts}       % blackboard math symbols
\usepackage{nicefrac}       % compact symbols for 1/2, etc.
\usepackage{microtype}      % microtypography
\usepackage{cleveref}       % smart cross-referencing
\usepackage{lipsum}         % Can be removed after putting your text content
\usepackage{graphicx}
\usepackage{natbib}
\usepackage{doi}

\usepackage{tabularx}   % flexible column widths (X)
\usepackage{enumitem}   % control list spacing/indent
\usepackage{multicol}   % multi-column lists
\usepackage{array}
\usepackage{multirow}

\usepackage{array,makecell}

\usepackage{array,multicol}
\newcolumntype{P}[1]{>{\raggedright\arraybackslash}p{#1}}
\newcolumntype{M}[1]{>{\centering\arraybackslash}m{#1}} % centered + vertically centered
\newcommand{\vcentered}[1]{\raisebox{2.2\height}{#1}}
\newcommand{\vcentereded}[1]{\raisebox{1\height}{#1}}

\usepackage{authblk}
\renewcommand{\arraystretch}{1.2} % a bit more row height

\title{AIvailable: A Software-Defined Architecture for LLM-as-a-Service on Heterogeneous and Legacy GPUs}

\newif\ifuniqueAffiliation
\uniqueAffiliationtrue

\usepackage{authblk}
\author[1]{Pedro Antunes}
\author[1]{Ana Rita Ortigoso\thanks{\href{https://orcid.org/0009-0001-7529-5857}{ORCID: 0009-0001-7529-5857}}}
\author[1]{Gabriel Vieira\thanks{\href{https://orcid.org/0009-0000-2300-8441}{ORCID: 0009-0000-2300-8441}}}
\author[1]{Daniel Fuentes\thanks{\href{https://orcid.org/0000-0001-9726-1087}{ORCID: 0000-0001-9726-1087}}}
\author[1]{Luís Frazão\thanks{\href{https://orcid.org/0000-0003-2571-7940}{ORCID: 0000-0003-2571-7940}}}
\author[1]{Nuno Costa\thanks{\href{https://orcid.org/0000-0002-2353-369X}{ORCID: 0000-0002-2353-369X}}}
\author[1]{António Pereira\thanks{\href{https://orcid.org/0000-0001-5062-1241}{ORCID: 0000-0001-5062-1241}}}

\affil[1]{Computer Science and Communication Research Centre, Polytechnic University of Leiria, Portugal\\
\texttt{(pedro.m.antunes, ana.l.ortigoso, gabriel.m.vieira, daniel.fuentes, luis.frazao, nuno.costa, apereira)@ipleiria.pt}}

\renewcommand{\shorttitle}{AIvailable}

\hypersetup{
pdftitle={AIvailable},
pdfsubject={cs.AI, cs.DC},
pdfauthor={Pedro Antunes, Ana Rita Ortigoso, Gabriel Vieira, Daniel Fuentes, Luís Frazão, Nuno Costa, António Pereira},
pdfkeywords={LLM, Heterogeneous, High Availability, Low-Cost, LLMaaS, SDAI},
}

\begin{document}
\maketitle

\begin{abstract}
	\noindent The rise of Large Language Models (LLM) has increased the need for scalable, high-performance inference systems, yet most existing frameworks assume homogeneous, resource-rich hardware, often unrealistic in academic, or resource-constrained settings. We introduce AIvailable, a low-cost, highly available LLM-as-a-Service (LLMaaS) platform, that uses a software-defined approach for running LLMs across heterogeneous and legacy GPU nodes, including NVIDIA and AMD devices, with a focus on fully utilizing each node’s VRAM. AIvailable operates as a fully GPU-accelerated inference without CPU fallbacks, featuring a unified client interface that allows seamless interaction with all deployed LLMs through a single logical unit.
    The architecture comprises four main components: the Client Interface for user access, the Service Frontend for secure request routing and load balancing, the SDAI Controller for orchestration, deployment, and monitoring, and the Service Backend of heterogeneous GPU nodes executing workloads. By abstracting GPU-specific details and providing dynamic, VRAM-aware allocation and reallocation of models, AIvailable ensures efficient use of resources and resilience against failures or workload fluctuations. Targeting academic labs, private companies, and other constrained organizations, it supports diverse open LLMs helping democratize generative AI through the repurposing of legacy GPUs.
\end{abstract}

% keywords can be removed

\section{Introduction}
LLMs are increasingly being integrated into various aspects of daily life and professional work \cite{raza_industrial_2025}. However, not all individuals or organizations possess the resources to deploy high-end LLM solutions, which often require access to powerful GPUs. In practice, many institutions, such as schools, universities, and small-to-medium enterprises (SMEs), rely on mid-range hardware like older NVIDIA RTX series or older AMD series GPUs, and in some cases, even legacy devices such as the NVIDIA GTX series \cite{yee2024ondevicellmssmeschallenges}. Beyond resource constraints, some organizations also prefer to host LLMs locally due to concerns over data privacy, compliance requirements, or the need for tight integration with existing development workflows. This further reinforces the demand for low-cost, locally deployable solutions that can operate effectively on heterogeneous and resource-limited infrastructures.

To address this gap, we introduce AIvailable, a low-cost, highly available LLM-as-a-Service (LLMaaS) platform designed specifically for SMEs and institutions with limited GPU capacity. AIvailable enables effective LLM deployment without the need for high-end infrastructure, lowering barriers to entry while maintaining accessibility and performance.

This paper introduces an architecture designed to maximize the utilization of available hardware resources, with a particular focus on leveraging the full VRAM capacity of each computational node. The approach enables users to deploy and operate LLMs in a manner that fully exploits the capabilities of their selected hardware, regardless of heterogeneity across nodes. In addition, the architecture will utilize a unified client interface through which users can seamlessly communicate with all LLM instances they have deployed, across all chosen nodes, without the need to manage separate endpoints or configurations. In doing so, AIvailable is committed to democratising the use of LLMs for everyone.

\section{Related Work}

The deployment of LLMs on heterogeneous and resource-constrained infrastructures has become a growing area of interest, particularly as organizations seek alternatives to high-end datacenter solutions. Several approaches have been proposed to address challenges related to availability, efficiency, and accessibility.

On this matter,  \cite{jiang2025thunderservehighperformancecostefficientllm} introduces a distributed serving architecture designed to provide low-latency inference across GPU clusters while maintaining resilience to node failures. Although effective in high-performance settings, this approach assumes access to datacenter-grade accelerators (e.g., A100 or H100 GPUs), which makes it less suitable for SMEs or educational institutions relying on legacy hardware.

Complementary to this, \cite{jha2024learnedbesteffortllmserving} propose adaptive scheduling policies for LLM serving, aiming to maximize GPU memory and compute efficiency under varying workloads. Their work demonstrates near-optimal VRAM utilization, but evaluations have largely been conducted on homogeneous and modern GPU clusters, leaving open questions regarding deployment on heterogeneous and resource-limited systems.

Task scheduling for distributed and heterogeneous infrastructures has also been explored in the Berkeley technical report \cite{Ren:EECS-2024-111}, which investigates decentralized strategies for allocating workloads across diverse nodes. While the framework highlights the potential of heterogeneous orchestration, it is designed under the assumption of stable high-bandwidth interconnects, which are often not available in smaller institutional settings.

More recently, research has also examined the role of LLMs in universities. The FLEXI Project \cite{10837635}, describes the establishment of an open LLM infrastructure. FLEXI demonstrates how locally maintained open-source models can support teaching and research while addressing issues of data protection, operating costs, and educational equity. By leveraging moderate hardware, the project enables experimentation with LLMs without relying on commercial providers, thereby giving universities greater control over privacy and compliance concerns.

Together, these efforts provide valuable foundations for building scalable and efficient LLM infrastructures. However, most existing systems either assume homogeneous high-performance hardware or focus on experimental proofs-of-concept within well-resourced institutions. In contrast, \textbf{AIvailable} specifically targets SMEs and academic environments by enabling reliable deployment on heterogeneous and legacy GPU devices, while offering a unified interface for high availability and efficient resource utilization.

\section{Proposed Architecture}

\begin{figure*}[!htpb]
    \centering
    \includegraphics[width=0.75\linewidth]{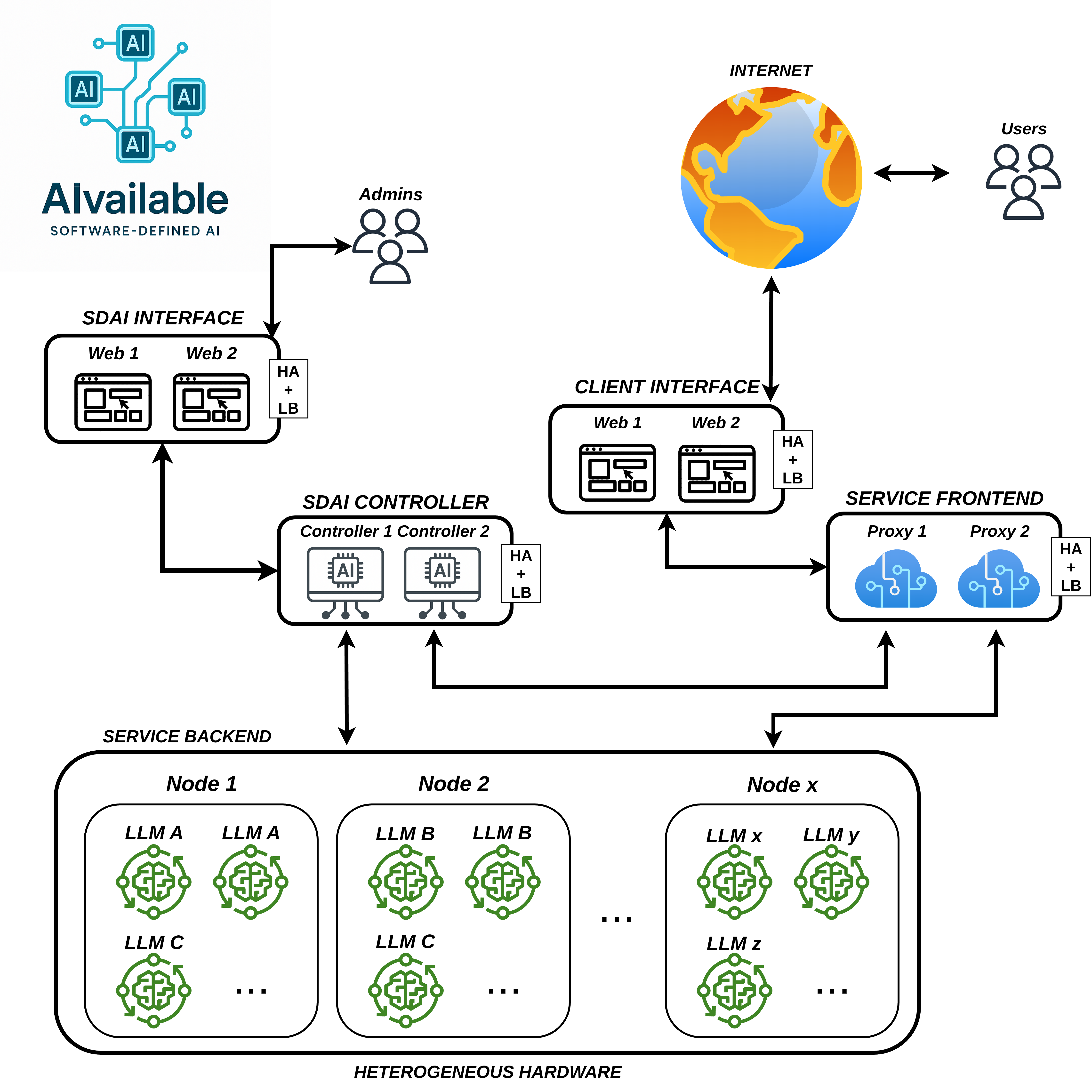}
    \caption{AIvailable Architecture}
    \label{fig:arch}
\end{figure*}

The proposed architecture, as depicted in Figure~\ref{fig:arch}, was designed to address the fundamental challenge of deploying and managing LLMs in environments with limited and heterogeneous GPU resources. Its purpose is to provide a low-cost yet reliable alternative to datacenter-scale infrastructures, ensuring that institutions such as SMEs and universities can still benefit from modern LLM capabilities. 
To achieve this, the architecture balances three core objectives: \textit{High Availability}, by mitigating single points of failure and enabling seamless failover, \textit{Efficient Resource Utilization}, by fully exploiting the VRAM capacity of each available device; and \textit{Ease of Access}, by exposing a unified and transparent interface to end-users regardless of the underlying hardware diversity. 

To support these objectives, the system is organized into four main components: the \textbf{Client Interface}, the \textbf{Service Frontend}, the \textbf{Software-Defined AI} (SDAI) \textbf{Controller}, and the \textbf{Service Backend}. Each component plays a distinct role in ensuring that LLMs can be deployed flexibly and operated sustainably across heterogeneous infrastructures.

The \textbf{Client Interface} serves as the primary environment for end-users to interact with the deployed LLMs. It provides access to models already running on the backend nodes, allowing users to send requests and receive responses without concern for the underlying routing or node selection. It does not handle model provisioning or deployment decisions, focusing solely on providing a streamlined and reliable point of interaction for inference.

The \textbf{Service Frontend} acts as the secure gateway for all client requests. It receives incoming interactions, routes them to the appropriate backend resources, and integrates HA and LB mechanisms to prevent node overload. The close integration between the \textbf{Service Frontend} and the \textbf{SDAI Controller} ensures that routing decisions are always based on the most up-to-date status of all nodes, preventing requests from being sent to unavailable or overloaded resources.

The \textbf{SDAI Controller} is the orchestration core of the system and includes its own dedicated \textbf{WebUI} for model selection and deployment management. Upon startup, it discovers and establishes communication with all backend nodes and the \textbf{Service Frontend}, registering their capabilities and current state. Through the \textbf{Controller’s WebUI}, administrators can select the nodes on which specific LLMs will run, manage configurations, and initiate deployments. Once models are deployed, the \textbf{Controller} provisions access via the \textbf{Service Frontend} and continuously monitors node health, updating the UI with relevant operational data.

The \textbf{Service Backend} is composed of heterogeneous computing nodes, ranging from modern mid-tier GPUs to legacy GPUs. These nodes execute the LLM workloads assigned by the \textbf{SDAI Controller}. The system supports multiple concurrent LLM instances per node and can dynamically reallocate models to different nodes in response to workload changes or hardware constraints, ensuring efficient use of VRAM.

Operationally, the system begins with the \textbf{SDAI Controller’s} discovery phase, during which it detects all available nodes, their capabilities, and any preloaded models. Administrators then use the \textbf{Controller’s WebUI} to select the target nodes and LLMs for deployment. Once deployed, the models become accessible to end-users via the \textbf{Client Interface} through the \textbf{Service Frontend}. Throughout operation, the \textbf{SDAI Controller} monitors utilization, network health, and model performance, dynamically reallocating workloads as necessary to maintain efficiency and service availability.

\section{Prototype}
\begin{figure*}[!htpb]
    \centering
    \includegraphics[width=0.75\linewidth]{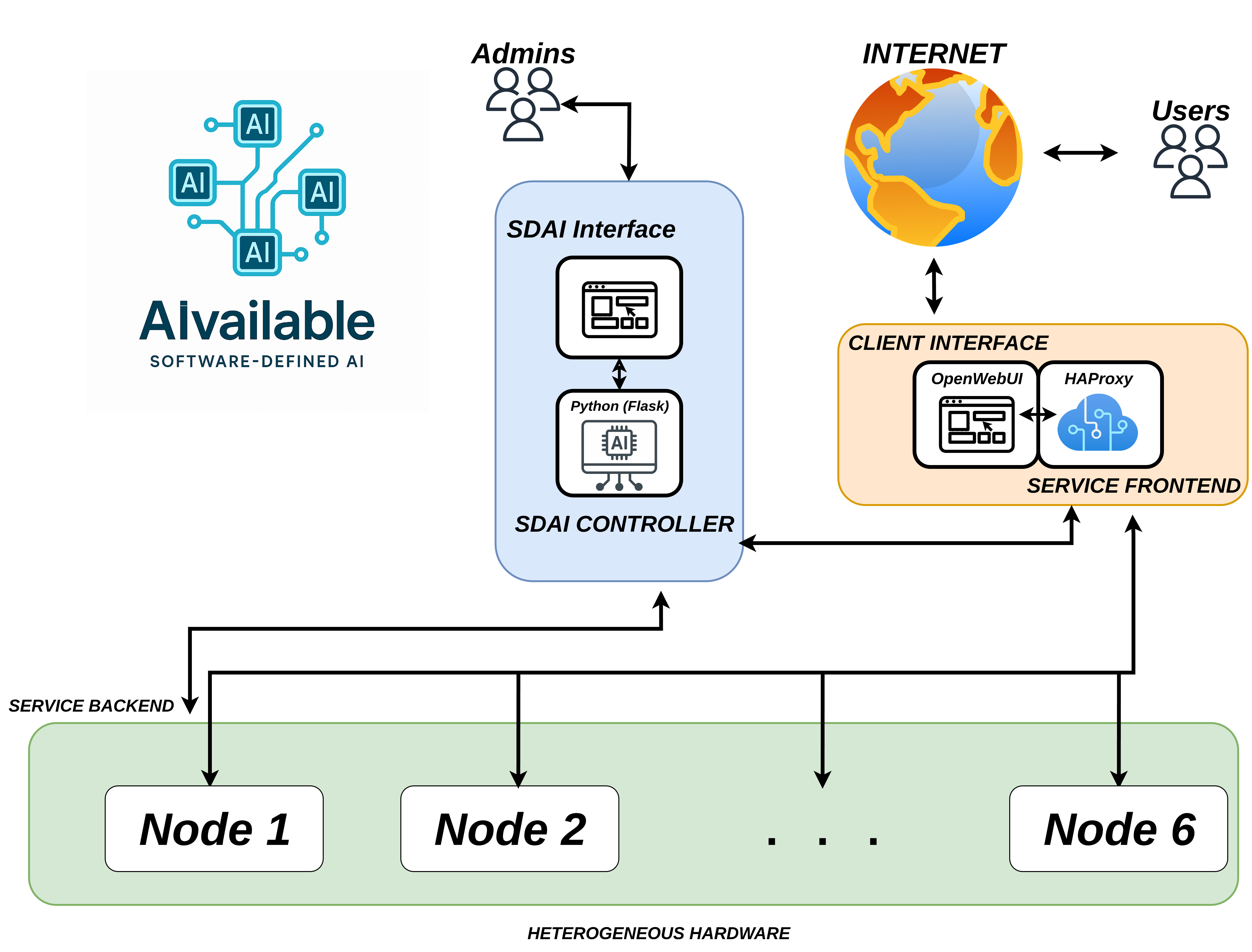}
    \caption{AIvailable Prototype}
    \label{fig:prototype}
\end{figure*}

In accordance with the proposed architecture, a prototype implementation was developed to validate the system’s feasibility, performance, and operational flow. The prototype follows the same architectural pattern, with all components deployed on heterogeneous hardware to reflect realistic constraints faced by SMEs and educational institutions. Figure~\ref{fig:prototype} shows the implemented prototype along with the hardware specifications of each machine used.

The \textbf{Client Interface} in the prototype is implemented using OpenWebUI, providing a lightweight yet feature-rich web-based environment for end-users to interact with the deployed LLMs. It is designed solely for inference usage, allowing users to send requests to models that are already running on the backend nodes. It does not handle model provisioning or deployment decisions, which are instead managed entirely through the \textbf{SDAI Controller’s WebUI}.

The \textbf{Service Frontend} employs HAProxy to provide load balancing and request routing, ensuring that inference queries are efficiently directed to the appropriate backend nodes while maintaining redundancy for HA. By leveraging HAProxy’s proven performance and flexibility, the system benefits from features such as health checking, connection pooling, and fine-grained traffic control. This not only maximizes throughput and fault tolerance across heterogeneous GPU backends but also simplifies scaling and operational management, making it a robust solution for production-grade LLM deployments

The \textbf{SDAI Controller}, implemented in Python using the Flask framework, serves as the orchestration core, managing node discovery, LLM allocation, and service lifecycle. Upon deployment, the controller not only provisions access through the \textbf{Service Frontend} but also sends each node a tailored HAProxy configuration based on the selected models, along with a startup script to launch the LLM instances using the Ollama framework.

The \textbf{Service Backend} in the prototype consists of six heterogeneous nodes with varying hardware capabilities, including configurations with NVIDIA RTX 3070, NVIDIA 1660 Super, AMD RX 6600 and AMD RX 6800. Each node is capable of hosting multiple LLMs matched to their available VRAM, were Models were chosen according to the computational resources available at each node, these being chosen in the SDAI Controller for each node. The list of the available LLM Models for each node, can be viewed on Table~\ref{tab:llm_nodes}.
Additionally, every backend node runs its own instance of HAProxy, enabling multiple replicas of the same model to run in the same node or across different nodes. This configuration improves scalability by distributing inference workloads across replicas and enhances resilience by allowing requests to be rerouted if a particular instance fails. 

\begin{table}[!htpb]
\centering
\caption{LLM Models deployed across Nodes}
\label{tab:llm_nodes}
\setlength{\tabcolsep}{6pt}
\renewcommand\arraystretch{1.15}
\begin{tabular}{|M{1.8cm}|P{14.5cm}|}
\hline
\textbf{Nodes} & \textbf{Models} \\ \hline

\vcentered{1, 2, 4} &
\begin{minipage}[t]{\linewidth}
\begin{multicols}{3}\raggedright
deepseek-r1: 1.5b, 7b, 8b\\
gemma3: 1b, 4b(v)\\
llama3.2: 1b, 3b\\
qwen2.5vl: 3b(v)\\
qwen3: 1.7b, 4b, 8b\\
mxbai-embed-large\\
nomic-embed-text\\
\end{multicols}
\end{minipage} \\ \hline

\vcentereded{3, 5} &
\begin{minipage}[t]{\linewidth}
\begin{multicols}{3}\raggedright
deepseek-r1: 1.5b, 7b\\
gemma3: 1b\\
llama3.2: 1b, 3b\\
qwen3: 1.7b, 4b\\
mxbai-embed-large\\
nomic-embed-text\\
\end{multicols}
\end{minipage} \\ \hline

\vcentered{6} &
\begin{minipage}[t]{\linewidth}
\begin{multicols}{3}\raggedright
deepseek-r1: 1.5b, 7b, 8b\\
gemma3: 1b, 4b(v)\\
qwen2.5vl: 3b(v)\\
llama3.2: 1b, 3b, 11b(v)\\
qwen3: 1.7b, 4b, 8b\\
mxbai-embed-large\\
nomic-embed-text\\
\end{multicols}
\end{minipage} \\ \hline

\end{tabular}
\end{table}

Table~\ref{tab:nodes_hardware} summarizes the hardware configuration of all backend nodes as well as the dedicated machine serving as the \textbf{Service Frontend}. These specifications highlight the heterogeneous nature of the deployment environment, ranging from modern mid-tier GPUs to legacy devices.

\begin{table}[!htpb]
\centering
\caption{Nodes Hardware Details}
\label{tab:nodes_hardware}
\begin{tabular}{|l|l|l|l|l|}
\hline
\textbf{Node} & \textbf{Storage} & \textbf{GPU}                       & \textbf{GPU Accel. Toolkit} & \textbf{GPU Year} \\ \hline
1             & SSD 120GB        & 1x AMD RX 6600 8GB GDDR6           & ROCm                        & 2021              \\ \hline
2             & SSD 240GB        & 1x NVIDIA RTX 3070 8GB GDDR6       & CUDA                        & 2020              \\ \hline
3             & SSD 120GB        & 1x NVIDIA GTX 1660 Super 6GB GDDR6 & CUDA                        & 2019              \\ \hline
4             & SSD 240GB        & 1x AMD RX 6600 8GB GDDR6           & ROCm                        & 2021              \\ \hline
5             & SSD 480GB        & 2x NVIDIA GTX 1660 Super 6GB GDDR6 & CUDA                        & 2019              \\ \hline
6             & NVME 250GB       & 1x AMD RX 6800 RX 16GB GDDR6       & ROCm                        & 2020              \\ \hline
\end{tabular}
\end{table}

\section{SDAI Interface}
This section provides an overview of the SDAI Interface, outlining its purpose, design, and core features. To enhance understanding, we include illustrative examples and images that demonstrate the interface in action, showcasing how it facilitates interaction and delivers value to end-users.

\begin{figure}[!htpb]
    \centering
    \includegraphics[width=1\textwidth]{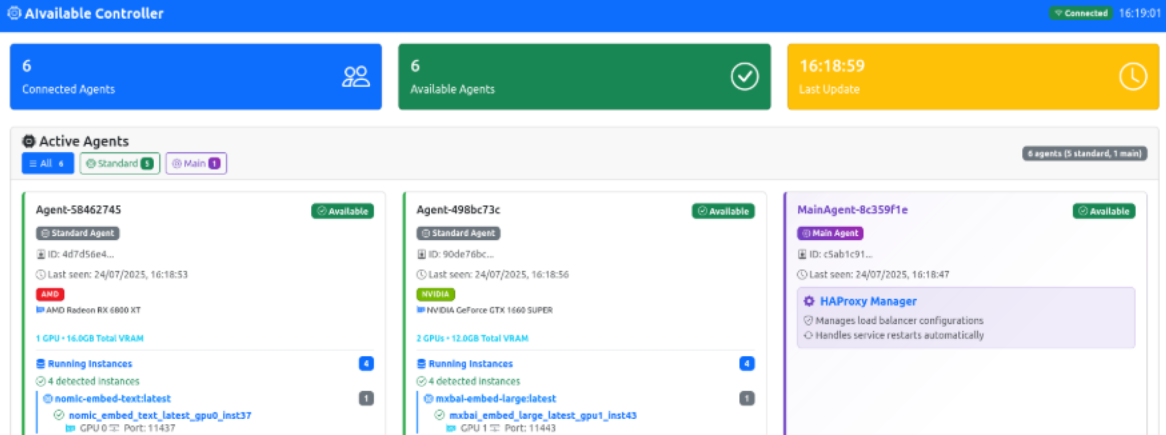}
    \caption{SDAI Interface Dashboard}
    \label{fig:sdai_interface}
\end{figure}

The first element presented in Figure~\ref{fig:sdai_interface} is the main dashboard of the \textbf{SDAI Interface}. 
This interface provides a comprehensive, real-time overview of all connected agents and their associated workloads, thereby serving as a central point for monitoring and management. The dashboard is composed of several key components.

The \textbf{Controller Overview}, located at the top bar, displays the number of connected and available agents, the last update timestamp, and the current connection status. 
The \textbf{Active Agents Section} presents each agent as a card, detailing hardware specifications such as GPU type, VRAM, and last communication time. 
It also lists the running model instances, which include both LLMs (e.g., \textit{llama3.2}, \textit{gemma3}, \textit{deepseek-r1}) and embedding models (e.g., \textit{mxbai-embed-large}, \textit{nomic-embed-text}). 
The \textbf{Main Agent}, highlighted in purple, hosts the HAProxy Manager, which manages load balancing and service restarts, ensuring high availability and reliability. Finally, at the bottom of the dashboard, the \textbf{Configuration Wizard} provides tools to generate optimized HAProxy and Ollama configurations based on GPU capabilities and model requirements. Together, these features make the interface a central hub for monitoring, managing, and scaling AI deployments.

\begin{figure}[!htpb]
    \centering
    \includegraphics[width=0.85\textwidth]{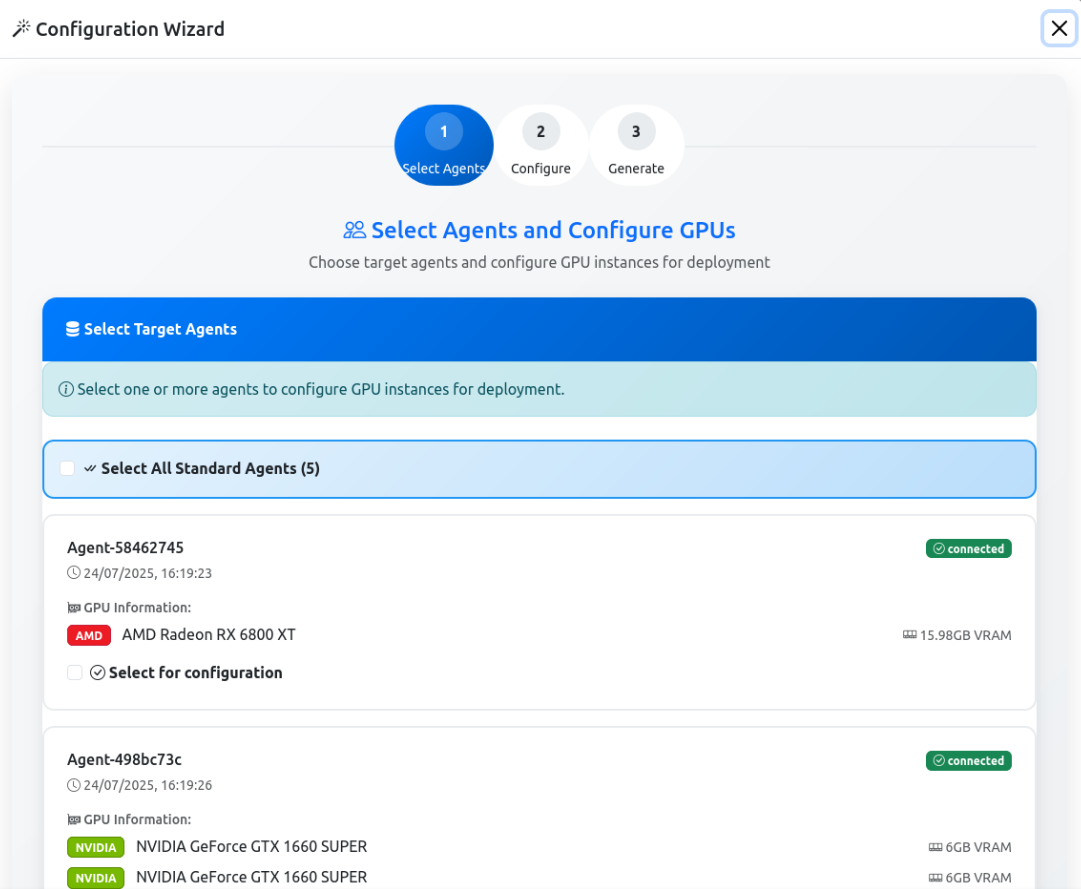}
    \caption{SDAI Configuration Wizard}
    \label{fig:sdai_interface_config}
\end{figure}

After that, we can view Figure~\ref{fig:sdai_interface_config}, which presents the \textbf{Configuration Wizard} of the SDAI Interface. This component enables users to select agents and configure GPU instances for deployment, thereby ensuring that resources are allocated according to system requirements. The wizard follows a stepwise approach that facilitates clarity and usability. It is divided into three stages: \textbf{Select Agents}, \textbf{Configure}, and \textbf{Generate}, which together guide the user through the configuration process in a systematic manner. During the \textbf{Target Agent Selection} stage, users may choose one or more agents for GPU deployment, with the option to select all standard agents simultaneously, simplifying larger deployments. For each agent, the interface provides detailed information, including its connection status, last communication timestamp, and GPU specifications such as vendor (e.g., AMD, NVIDIA), model, and available VRAM. In addition, users are given an explicit option to include or exclude each agent from the configuration process.

\clearpage

\subsection{Select Agents}
\begin{figure}[!htbp]
    \centering
    \includegraphics[width=0.60\textwidth]{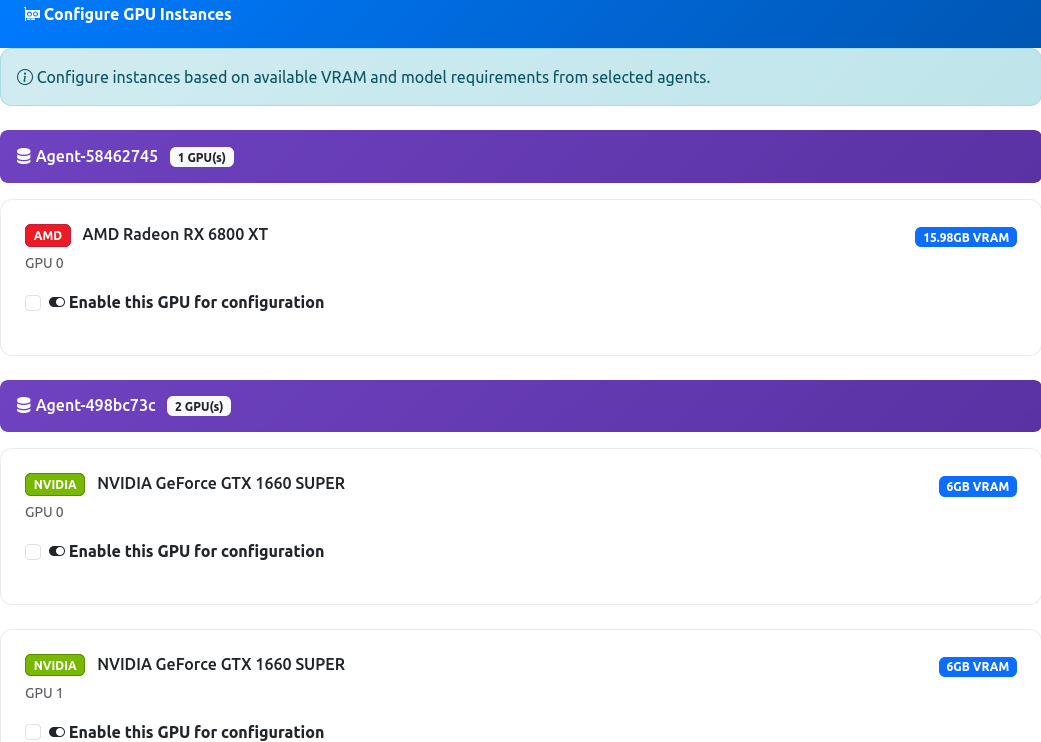}
    \caption{SDAI GPU Selection}
    \label{fig:sdai_interface_agents}
\end{figure}

A more detailed view of the \textit{Select Agents} stage is provided in Figure~\ref{fig:sdai_interface_agents}. Once an agent has been selected, the user is presented with the option to enable specific \textbf{GPU instances} for configuration, showing a clear listing of available GPUs for each selected agent, model type, and associated VRAM capacity. There is also an enable/disable toggle for each GPU, allowing the user to select precisely which devices will participate in the deployment process.

\begin{figure}[!htbp]
    \centering
    \includegraphics[width=0.60\textwidth]{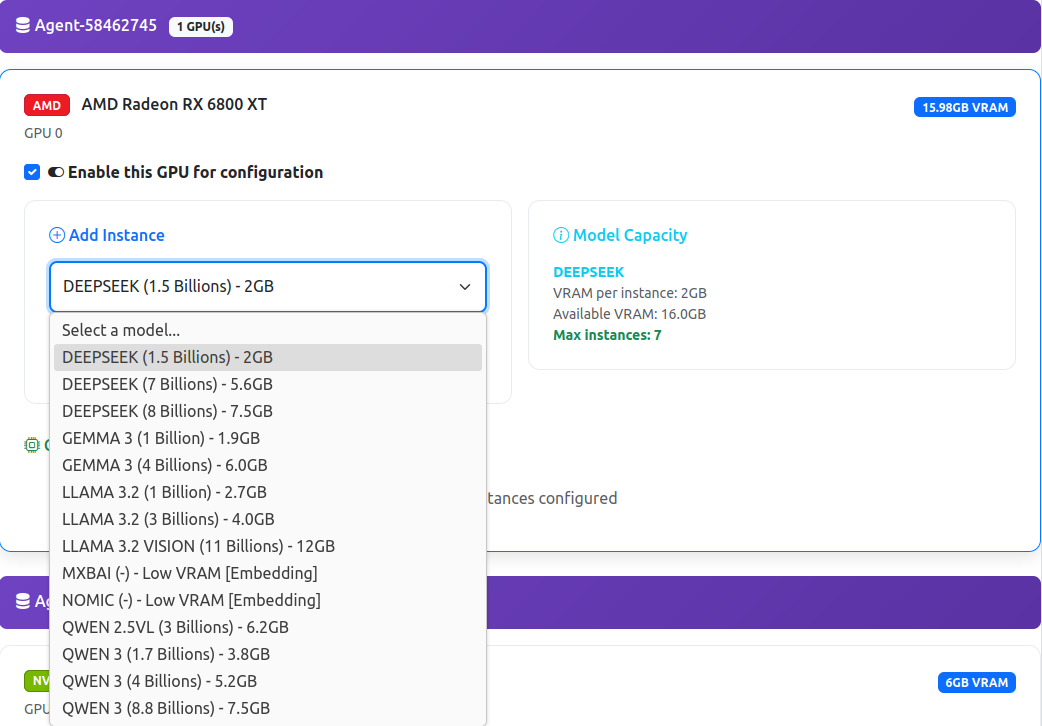}
    \caption{SDAI Model Selection}
    \label{fig:sdai_interface_models}
\end{figure}

After enabling a GPU for configuration, the user is provided with the option to select which \textbf{LLM instances} will be deployed. 
As shown in Figure~\ref{fig:sdai_interface_models}, this selection is made from a predefined list of available models, which includes a variety of architectures and parameter sizes (e.g., \textit{DeepSeek}, \textit{Gemma}, \textit{Llama}, \textit{Qwen}, and embedding models). 
The interface also displays information about \textbf{model capacity}, including the VRAM required per instance, the available VRAM on the selected GPU, and the maximum number of instances that can be allocated. 

\clearpage

This design enables users to align model selection with hardware constraints, ensuring efficient resource utilization while maintaining flexibility in the deployment of different model families and sizes.

\subsection{Configure}
\begin{figure}[!htbp]
    \centering
    \includegraphics[width=0.85\textwidth]{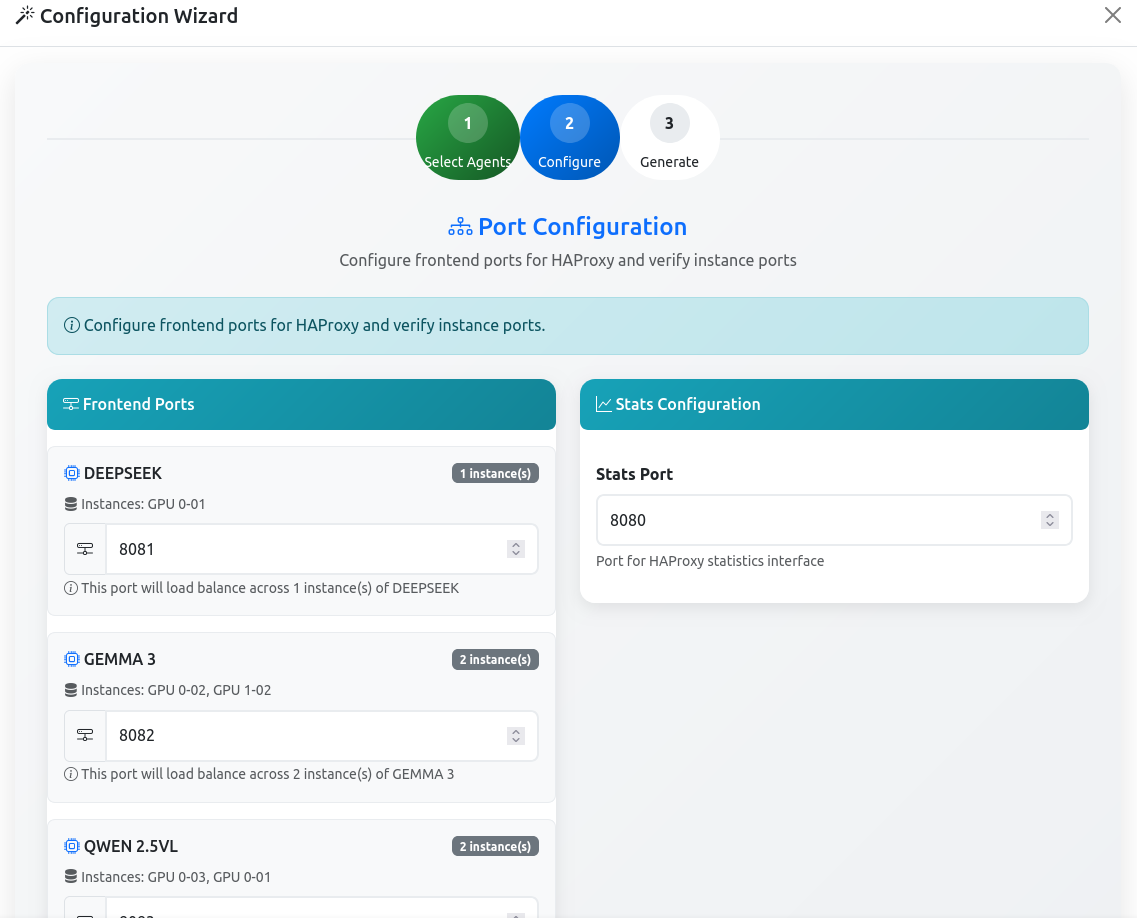}
    \caption{SDAI Port Configuration}
    \label{fig:sdai_interface_ports}
\end{figure}

After selecting the agents and assigning specific model instances to each GPU, the process advances to the \textbf{Configure} stage. 
As illustrated in Figure~\ref{fig:sdai_interface_ports}, this stage allows the user to define the \textbf{network ports} through which each model instance will be accessed. 
For each deployed model (e.g., Qwen 2.5VL, DeepSeek, Qwen 3, Nomic, MXBAI), the interface provides:

\begin{itemize}
    \item A clear indication of the number of running instances and the GPUs on which they are hosted.
    \item An automatically suggested default port, with the option for the user to manually adjust it as required.
    \item A load balancing mechanism across instances, ensuring that requests are evenly distributed for models with multiple replicas.
\end{itemize}

\subsection{Generate}
\begin{figure}[!htbp]
    \centering
    \includegraphics[width=0.80\textwidth]{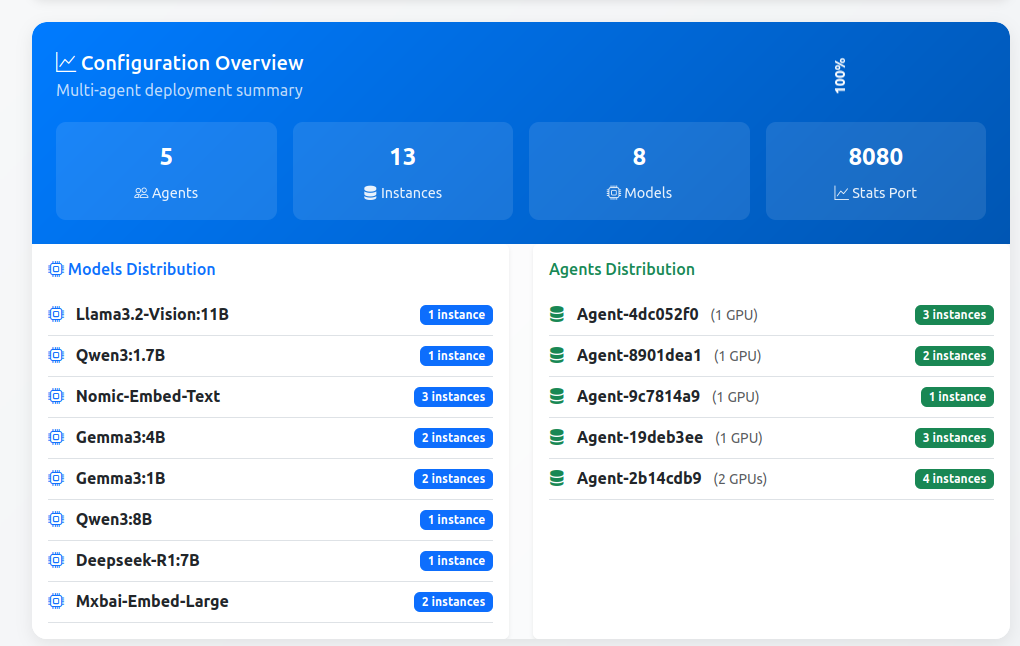}
    \caption{SDAI Configuration Overview}
    \label{fig:sdai_interface_overview}
\end{figure}

The final stage, \textbf{Generate}, presents the \textbf{Configuration Overview}, which provides a comprehensive summary of the multi-agent deployment. 
As shown in Figure~\ref{fig:sdai_interface_overview}, this stage consolidates the configuration details and displays them in a structured format, enabling users to verify the allocation of resources before finalizing the setup. The overview includes:

\begin{itemize}
    \item \textbf{System Statistics} - a summary panel indicating the total number of agents, deployed instances, distinct models, and the statistics port being used.
    \item \textbf{Model Distribution} - a breakdown of all deployed models, along with the number of instances allocated to each.
    \item \textbf{Agent Distribution} - a detailed mapping of how many instances are deployed per agent, as well as the GPU resources available to each agent.
\end{itemize}

\section{Discussion}

\textbf{AIvailable} demonstrates that a software-defined approach can make multi-agent LLM serving feasible on heterogeneous and partially legacy GPU fleets.
The SDAI Interface reduced orchestration complexity by exposing a single control surface for agent discovery, model placement, port assignment, and HA configuration. HAProxy’s role simplified rollout and rollback, and its health checks provided early detection for instance drift.

Replica-level load balancing masked single-instance failures; however, it does not address \emph{coordinated} failures (e.g., node-level outages). Observability currently focuses on liveness and capacity rather than application-level quality.

The stepwise wizard (\textbf{Select~$\rightarrow$~Configure~$\rightarrow$~Generate}) lowered the barrier for non-experts, but also centralized power. Furthermore, local hosting was a motivator for privacy; documenting data-handling guarantees will be important for institutional adoption.

We did not provide quantitative benchmarks in this study. As such, claims about agility and responsiveness remain qualitative. External validity is limited by the size of our testbed and by the particular set of open models considered. Finally, cost analysis (energy, amortized hardware, and ops) was out of scope.

% Conclusion
\section{Conclusion}

This work introduced \textbf{AIvailable}, a software-defined architecture and working prototype for LLM-as-a-Service on heterogeneous and legacy GPU infrastructures. Our main contributions are:
\begin{enumerate}
    \item An architecture that couples a lightweight control plane (SDAI Controller) with a standards-based data plane (HAProxy), enabling high availability and unified access across diverse nodes.
    \item A practical orchestration workflow, delivered through the SDAI Interface, that streamlines agent discovery, VRAM-aware model placement, and configuration generation.
    \item A demonstration on mixed ROCm/CUDA hardware showing that reliable, multi-agent LLM serving is attainable without datacenter-class GPUs.
\end{enumerate}

Looking ahead, the developed architecture and prototype set the foundation for further advancements in orchestrating modern LLMs on non-uniform, GPU-driven infrastructures without reliance on high-end datacenter hardware. Future work will involve large-scale validation, including benchmarking inference latency, throughput across diverse model configurations, and assessing scalability under realistic workloads. Additional directions include exploring advanced strategies for dynamic model allocation and investigating mechanisms to further enhance fault tolerance. With these steps, we aim for \textbf{AIvailable} to contribute meaningfully to democratizing access to generative AI in academic, research, and SME environments.

\bibliographystyle{unsrt}
\bibliography{references}

@article{raza_industrial_2025,
	title = {Industrial applications of large language models},
	volume = {15},
	issn = {2045-2322},
	url = {https://www.nature.com/articles/s41598-025-98483-1},
	doi = {10.1038/s41598-025-98483-1},
	language = {en},
	number = {1},
	urldate = {2025-08-15},
	journal = {Scientific Reports},
	author = {Raza, Mubashar and Jahangir, Zarmina and Riaz, Muhammad Bilal and Saeed, Muhammad Jasim and Sattar, Muhammad Awais},
	month = apr,
	year = {2025},
	pages = {13755},
}

@misc{yee2024ondevicellmssmeschallenges,
      title={On-Device LLMs for SMEs: Challenges and Opportunities}, 
      author={Jeremy Stephen Gabriel Yee and Pai Chet Ng and Zhengkui Wang and Ian McLoughlin and Aik Beng Ng and Simon See},
      year={2024},
      eprint={2410.16070},
      archivePrefix={arXiv},
      primaryClass={cs.AI},
      url={https://arxiv.org/abs/2410.16070}, 
}

@mastersthesis{Ren:EECS-2024-111,
    Author= {Ren, Elden},
    Title= {Task Scheduling for Decentralized LLM Serving in Heterogeneous Networks},
    School= {EECS Department, University of California, Berkeley},
    Year= {2024},
    Month= {May},
    Url= {http://www2.eecs.berkeley.edu/Pubs/TechRpts/2024/EECS-2024-111.html},
    Number= {UCB/EECS-2024-111},
}

@misc{jha2024learnedbesteffortllmserving,
      title={Learned Best-Effort LLM Serving}, 
      author={Siddharth Jha and Coleman Hooper and Xiaoxuan Liu and Sehoon Kim and Kurt Keutzer},
      year={2024},
      eprint={2401.07886},
      archivePrefix={arXiv},
      primaryClass={cs.LG},
      url={https://arxiv.org/abs/2401.07886}, 
}

@misc{jiang2025thunderservehighperformancecostefficientllm,
      title={ThunderServe: High-performance and Cost-efficient LLM Serving in Cloud Environments}, 
      author={Youhe Jiang and Fangcheng Fu and Xiaozhe Yao and Taiyi Wang and Bin Cui and Ana Klimovic and Eiko Yoneki},
      year={2025},
      eprint={2502.09334},
      archivePrefix={arXiv},
      primaryClass={cs.DC},
      url={https://arxiv.org/abs/2502.09334}, 
}

@INPROCEEDINGS{10837635,
  author={Zesch, Torsten and Hanses, Michael and Seidel, Niels and Aggarwal, Piush and Veiel, Dirk and De Witt, Claudia},
  booktitle={2024 21st International Conference on Information Technology Based Higher Education and Training (ITHET)}, 
  title={Flexible LLM Experimental Infrastructure (Flexi) – Enabling Experimentation and Innovation in Higher Education Through Access to Open LLMs}, 
  year={2024},
  volume={},
  number={},
  pages={1-8},
  doi={10.1109/ITHET61869.2024.10837635}}

\end{document}